# Ultrafast switching time and third order nonlinear coefficients of microwave treated single walled carbon nanotube suspensions

N Kamaraju[1], Sunil Kumar[1], B. Karthikeyan[1], Bhalchandra Kakade[2], Vijayamohanan K. Pillai[2] and A. K. Sood[1,*]

[1]Department of Physics and Centre for Ultrafast Laser Applications, Indian Institute of Science, Bangalore-560012, India.
[2]Physical and Materials chemistry Division, National Chemical Laboratory, Pune-411008, India.

Microwave treated water soluble and amide functionalized single walled carbon nanotubes have been investigated using femtosecond degenerate pump-probe and nonlinear transmission experiments. The time resolved differential transmission using 75 femtosecond pulse with the central wavelength of 790 nm shows a bi-exponential ultrafast photo-bleaching with time constants of of 160 fs (130 fs) and 920 fs (300 fs) for water soluble (amide functionalized) nanotubes. Open and closed aperture z-scans show saturation absorption and positive (negative) nonlinear refraction for water soluble (amide functionalized) nanotubes. Two photon absorption coefficient, $b_0 \sim$ 250 cm/GW (650 cm/GW) and nonlinear index, $\gamma \sim$ 15 cm$^2$/pW (-30 cm$^2$/pW) are obtained from the theoretical fit in the saturation limit to the data for two types of nanotubes.

* The corresponding author: asood@physics.iisc.ernet.in

7/9/2008 6:29:58 PM

## 1. Introduction

Single walled carbon nanotubes (SWNT) have evoked enormous interest due to their fascinating physical properties with many possible applications such as flow transducers[1], nanoelectronics[2,3] and nonlinear optics [4-6]. Quantum confinement of π-electrons[7] along the tube axis makes nanotubes one of the most promising materials with a large and ultrafast electronic third-order nonlinearity to be used in applications in terahertz optical switching for next generation ultrafast communication systems and passive mode-locking for the generation of femtosecond pulses. Thus the investigations of the nonlinearity and optical switching of SWNT using femtosecond pulses garner recent interests[6-14]. Unlike the nanosecond regime, in the femtosecond regime where heating does not play a role in the nonlinear transmission, an enormous large third order susceptibility (Im $\chi^{(3)}$ ~ $10^{-6}$ esu) has been reported by resonantly exciting at the first inter-subband energy level, $S_{11}$ of semi conducting SWNT[8,9]. Our recent femtosecond nonlinear transmission[10] studies on SWNT suspended in water at 1.57eV (non-resonant with $S_{11}$ and $S_{22}$) showed saturation absorption in open aperture (OA) and negative nonlinearity in closed aperture (CA) z-scan. In this article, we report both pump-probe and nonlinear transmission studies performed on water soluble and amide functionalized SWNT prepared using a microwave treatment. Absorption and Raman spectra are recorded to characterize the samples. The two photon absorption (TPA) coefficient, in this case is one order less compared to SWNT suspended in water containing sodium dodecyl sulphate (SDS)[10], whereas the nonlinear index is of the same order.

## 2. Sample Preparation

About 20 mg of SWNT (diameter 0.7 to 1.5 nm) was added to an 1:1 mixture of 98% $H_2SO_4$ and 78% $HNO_3$ and the mixture was subjected to microwave (MW) irradiation for 4 minutes using a domestic microwave oven using 60% of its total power of 700W [15-17]. The rigorous and harsh conditions imposed by the MW render rapid breaking of graphitic C=C to develop large amount of -OH, -COO- and –$SO_3H$ groups on the sidewalls of the SWNT as confirmed by FTIR studies. The soluble residue was first filtered through a 200 nm PTFE membrane and separated out from the unfiltered carbon matrix. The unfiltered residue was then dispersed in 3 M HCL and sonicated for 30 mins to develop –COOH groups rather than –COO-, which may hinder further functionalization. The acidified nanotubes were subjected to dialysis for 2 days to remove excess of acid content to make the dispersion neutral. The final product showed a typical high solubility of 2.6 mg/ml in water. However, it can be noted from the absorption spectra (shown in Fig. 1) that, the nanotubes are not severely damaged by microwave/acid treatment.

About 40 mg of sol- SWNT was ultrasonicated for 10 min in presence of 10 mL of oxalyl chloride in an argon atmosphere. The mixture was then refluxed for 24 hrs at 60 $^o$C to obtain a higher degree of acylchloride functionalization (SWNT-COCl); the excess of oxalyl chloride was subsequently removed under vacuum using rota-evaporation. SWNT-COCl was then mixed with 1 g solid tridecylamine ($CH_3(CH_2)_{12}NH_2$) and refluxed at 40 $^o$C (m. p. of TDA: ~30 $^o$C) for 40 hrs[18-20]. Excess of tridecylamine was then removed by washing with copious amount of THF with successive sonication and centrifugation. Solid amino derivatized sample of SWNT was collected after drying in air and characterized.

We call the soluble nanotubes in water as sol-SWNT and amide functionalized nanotubes in formamide (DMF) as amide-SWNT.

## 3. Experimental data

Optical absorption spectra of the samples were recorded using Bruker FT-IR absorption spectrometer. Raman spectra were recorded using a Dilor XY micro-Raman spectrometer equipped with a liquid nitrogen cooled CCD detector. The output from Ti:Sapphire Regenerative femtosecond amplifier (50 fs, 1.57 eV, 1 KHz Spitfire, Spectra Physics) was used for both z-scan and the degenerate pump-probe experiments. The pulse from the amplifier was found to be broadened to 80 fs near the sample point in z-scan experiments. Two Si-PIN diodes (one for the signal (B) and the other for reference (A)) triggered from the electronic clock output (1 kHz) from the amplifier are used for the data acquisition and the difference between B and A was collected using a lock-in amplifier (SRS 830), averaged over 300 shots. This difference data was then converted into actual B/A signal in a personal computer. The SWNT dispersion in 1 mm thick cuvette was translated using a motorized translation stage (XPS Motion controller, Newport) over the focal region. The intensity of input beam was varied from 150 MW/cm$^2$ to 12 GW/cm$^2$. In the OA z-scan, all the light was collected by using a collection lens in front of the diode. For CA z-scan, we kept an aperture of radius 3.6 mm in front of the diode B.

In the case of pump-probe experiments, the cross-correlation of pump and probe pulses was measured to be 75 fs (FWHM) using a thin BBO crystal at the sample point. The pump pulse was delayed in time using the computer controlled motorized translation stage. The change in the probe transmission due to the presence of the pump was monitored using two Si-PIN diodes (one for the reference beam and the other for the probe beam interacting with the pump) with the standard lock-in detection (pump beam was chopped at 139 Hz). The pump and probe intensities were kept at 1.5 GW/cm$^2$ and 58 MW/cm$^2$, respectively.

**4. Results and Discussion**

**(a) Absorption and Raman Spectra**

Fig. 1 shows the absorption spectra of sol-SWNT in water and amide-SWNT. The chosen photon energy of 1.57 eV lies in between the first interband transition energy of metallic tubes, $M_{11}$ and the second interband transition energy of the semi-conducting tubes $S_{22}$. The absorption coefficient, $\alpha_0$ at 1.57 eV is about 2.2 x 10$^4$ cm$^{-1}$ (3.6 x 10$^4$ cm$^{-1}$) for sol-SWNT (amide-SWNT). We also observe an unusually high absorption at 2.2 eV in sol-SWNT(Fig 1(a)) which matches with $S_{33}$, the third interband transition energy of the semi conducting tubes (diameter of 1.5 nm inferred from radial breathing mode at 170cm$^{-1}$ (see inset of Fig. 2 (a))) and this is blue shifted by 0.2 eV for amide-SWNT (Fig 1(b)). Another feature that is seen is the disappearance of the van Hove singularities ($S_{22}$ and $M_{11}$) in the case of amide functionalized SWNT, indicating the broadening of van Hove singularities in the electronic structure due to the production of defect sites on the sidewalls.

Fig. 2 shows Raman spectra of both the samples. The tangential mode is fitted with four Lorentzians at 1552cm$^{-1}$, 1567 cm$^{-1}$, 1590 cm$^{-1}$ and 1598 cm$^{-1}$ for sol-SWNT and 1566 cm$^{-1}$, 1586 cm$^{-1}$ 1599 cm$^{-1}$ and 1620 cm$^{-1}$ for amide-SWNT. The band near 1345 cm$^{-1}$ is the defect-activated D-mode. The ratio of the intensity of the D-mode to the tangential (T) mode at ~ 1590 cm$^{-1}$ ($I_D/I_T$) is 0.1 for sol SWNT and 0.5 for amide-SWNT, indicating large disorder in the case of amide-SWNT. The 2-D mode is at 2676 cm$^{-1}$ (2690 cm$^{-1}$) for sol (amide) -SWNT. We note that the T-mode frequencies in amide-SWNT are higher than in sol-SWNT. This can be due to hole doping of the nanotubes due to amide functionalization. In a hole doped Graphite Intercalation Compounds (GICs)[21,22], [$\Delta\omega/f_C$] ~ 460 cm$^{-1}$ where $\Delta\omega$ is peak shift of the doped sample with respect to the undoped sample and $f_C$ is the degree of charge transfer between carbon atoms and the dopants. Applying this model to our case, we have $\Delta\omega = 9$ cm$^{-1}$ corresponding to $f_C$ ~ 2/100 (two holes per 100 carbon atoms in SWNT) for amide-SWNT.

**(b) Open Aperture Z-scan and closed Aperture Z-scan**

Fig. 3 (a) and (b) show the OA and CA z-scan profiles for the sol-SWNT. The results for the amide-SWNT are presented in Fig. 3(c) and (d). It can be seen that the OA z-scan for both the samples show saturable absorption. However, the CA z-scan reveals that the nonlinearity is of opposite sign: positive nonlinearity for sol-SWNT and negative nonlinearity for amide-SWNT.

In the limit of saturable absorption[10], we have the rate equation:

$$\frac{d\,I(z,\zeta,r,t)}{d\,x} = -\frac{\alpha_0\,I(z,\zeta,r,t)}{\left[1+\frac{I(z,\zeta,r,t)}{I_s}\right]} - \beta_{\text{eff}}\,I^2(z,\zeta,r,t); \quad (1)$$

$$\text{where}\quad \beta_{\text{eff}} = \left[\beta_0 + \frac{\sigma_0\alpha_0 t_0}{\hbar\omega}\right] \quad (2)$$

where $\zeta$ denotes the position in the sample, z is location of the sample, $\alpha_0$ is the one-photon absorption (including intrinsic free carrier absorption) coefficient, $\beta_0$ is the fundamental two photon absorption (TPA) coefficient, and $\sigma_0$ is free carrier absorption (FCA) cross section. $I_s$ is the parameter that characterizes the saturation absorption. The boundary condition required to solve equation (1) is the input intensity which is assumed to be Gaussian:

$$I(z,0,r,t) = I_0 \left[\frac{w_0}{w(z)}\right]^2 \exp\left[-\frac{2r^2}{w^2(z)}\right]\exp\left[-\frac{t^2}{t_0^2}\right] \quad (3)$$

Here, $w_0$ is the beam waist at the focus, $w(z) = w_0[1+(z/z_0)^2]^{1/2}$ is the beam radius at z, $z_0 = \frac{\pi w_0^2}{\lambda}$ is the diffraction length of the beam, $\tau_0$ is the half width at $e^{-1}$ of the maximum (HWe$^{-1}$M) of the pulse, and $\lambda$ is the wavelength. The intensity at the exit side of sample, I(z,L,r,t) is obtained from the analytical solution of equation (1) and integrated over all r and t to calculate the transmitted energy. The transmission in the OA z-scan experiment, $T_{OA}(z)$ is simply the ratio of transmitted energy to the incident energy. In the case of CA z-scan, we solve for the phase at the exit surface and then construct the appropriate electric filed from which we get[10] normalized z-scan transmittance, $T_{CA}(z)$. We have fitted both OA and CA z-scan data with the expressions for the normalized transmission (shown as solid lines in fig. 3). The consistent set of values for the parameters, $\beta_0$, $I_s$ and $\gamma$ that fit both OA and CA case are given in Table 1.

The reason for the change of sign in the nonlinear refractive index of sol-SWNT may be understood from the theoretical modeling of nonlinear refraction of nanotubes by Margulis and Guidik[7]. This model is based on the virtual excitation of $\pi$-electrons induced nonlinearity. The nonlinear refraction changes from negative to positive as the ratio of the photon energy to the band gap, $E_g$ in the semiconductor nanotubes, $\hbar \omega / E_g$ crosses 0.5. In our case, the relevant band gap for possible TPA is $E_{44}^S$ which is ~ 2.85 eV for sol-SWNT and ~ 3.2 eV for amide-SWNT. Therefore, $\hbar \omega / E_g$ ~ 0.54 for sol-SWNT and ~ 0.49 for amide-SWNT, thus explaining[7] the different signs of $\gamma$. In our earlier studies[10] on SDS suspension of nanotubes of diameter 1.4nm, $E_{44}^S$ ~ 3.2 eV thereby $\hbar \omega / E_g$ ~ 0.49 eV for which $\gamma$ should be negative as observed.

**(c) Open Degenerate pump-probe experiment**

In order to understand the recovery time of the observed large nonlinearities and to acquire an insight into the underlying mechanism, a degenerate pump-probe experiment has been carried out. Fig. 4 shows the normalized differential transmittance recorded with change in the time delay between pump and probe pulses. The observed data shown in fig. 4 could be very well fitted with the convolution of observed cross-correlation of an input pulse of 75 fs and a bi-exponential decay function. The ultrafast decay rates obtained are 160 fs (130 fs) and 920 fs (300 fs) for sol (amide)-SWNT. The amplitude of fast and slow components are 0.74 (0.85) and 0.26 (0.15). The observed fast and slow components of sol (amide)-SWNT here seem to agree well with the 130 fs and 1ps observed earlier[11-14] for SWNT suspension, which are attributed to intra-band carrier relaxation and interband carrier recombination respectively. These studies show that the fast component is always present in both the resonant and non-resonant excitations of the sample whereas the slow component strength increases in resonant excitation. In our case we see that the fast component of 160 (130 fs) is not affected whereas the slow component of the decay time 960 fs changed to 300 fs in amide-SWNT. The shorter decay time is attributed to the large number of defects in the amide-SWNT as seen in our Raman studies.

## 5. Conclusions:

We have investigated water soluble pristine SWNT and amide functionalized SWNT using femtosecond degenerate pump-probe and z-scan experiments. Large third order nonlinearity parameters [$b_0$ ~ 250 cm/GW (650 cm/GW) and $\gamma$ ~ 15 cm$^2$/pW (-30 cm$^2$/pW) for sol (amide)-SWNT] were estimated from femtosecond z-scan experiments. The ultrafast decay times (160 fs and 920 fs for sol-SWNT; 130 fs and 300 fs for amide-SWNT) are observed under non-resonant excitation of the nanotubes.

## 6. Acknowledgement:

AKS thanks Department of Science and Technology (DST), India for financial support.

**Table1:** The consistent set of values for the parameters, $\beta_0$, $I_s$ and $\gamma$ that fit both OA and CA case are given for sol-SWNT and Amide-SWNT.

|  | $\beta_0$ (cm/W) | $I_s$ (GW/cm$^2$) | $\gamma$ (cm$^2$/W) |
|---|---|---|---|
| Sol-SWNT | $(2.5 \pm 0.15) \times 10^{-7}$ | 34 | $(1.5 \pm 0.2) \times 10^{-11}$ |
| Amide-SWNT | $(6 \pm 0.15) \times 10^{-7}$ | 33 | $(-3 \pm 0.2) \times 10^{-11}$ |
| SWNT-SDS suspension[10] | $(14 \pm 0.15) \times 10^{-7}$ | 33 | $(-5.5 \pm 0.2) \times 10^{-11}$ |

**Figures:**

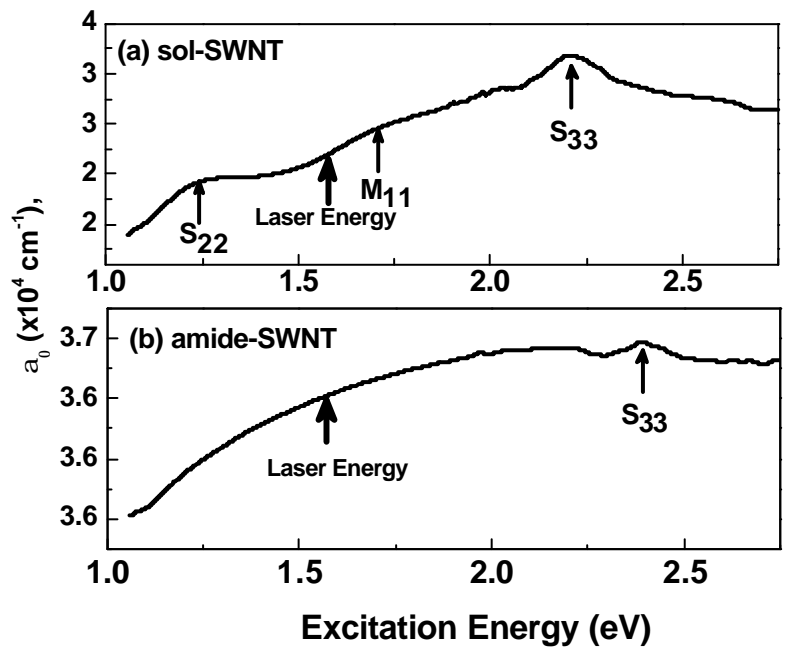

Fig. 1: Optical absorption spectra of (a) sol-SWNT and (b) microwave treated amide functionalized single walled carbon nanotubes in DMF. $S_{22}$, $S_{33}$ ($M_{11}$) are the second, third (first) interband energy of semi conducting (metallic) nanotubes. The arrow indicates the laser energy used.

Kamaraju et al.

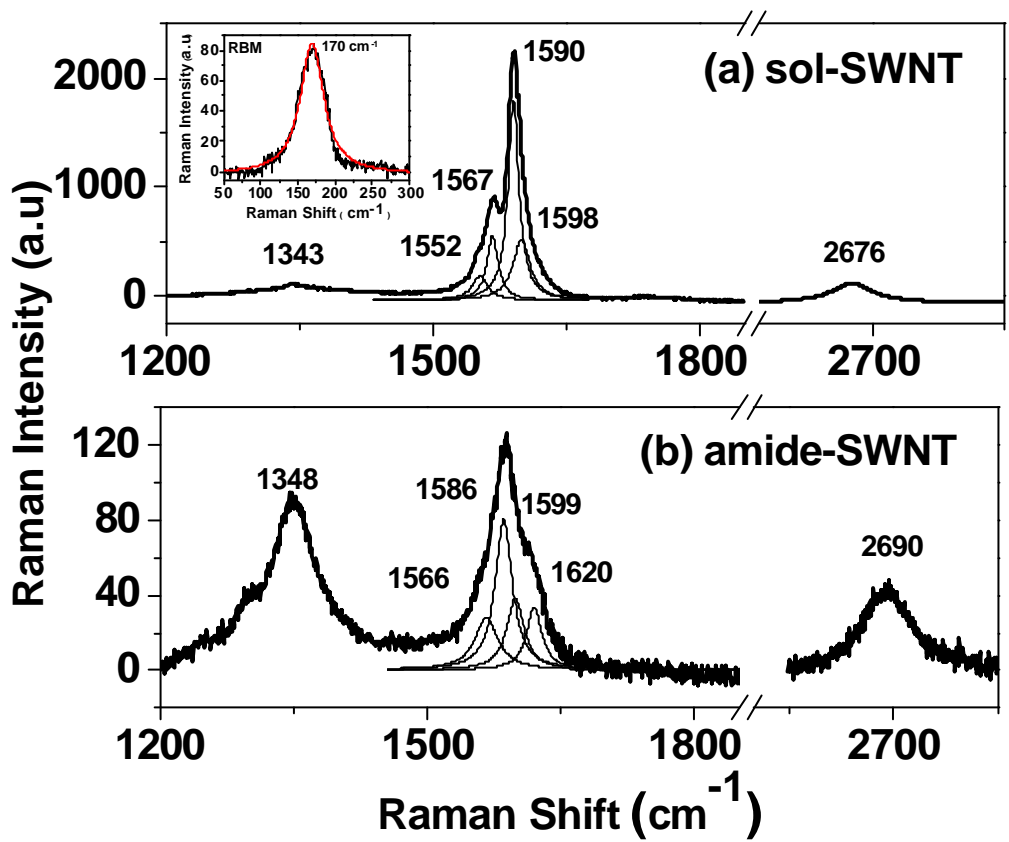

Fig. 2: Raman spectra (a) sol-SWNT (a) microwave treated amide functionalized single walled carbon nanotubes in DMF. The Lorentzians fitted to tangential modes are also shown. The inset in (a) is the radial breathing mode for sol-SWNT.

Kamaraju et al.

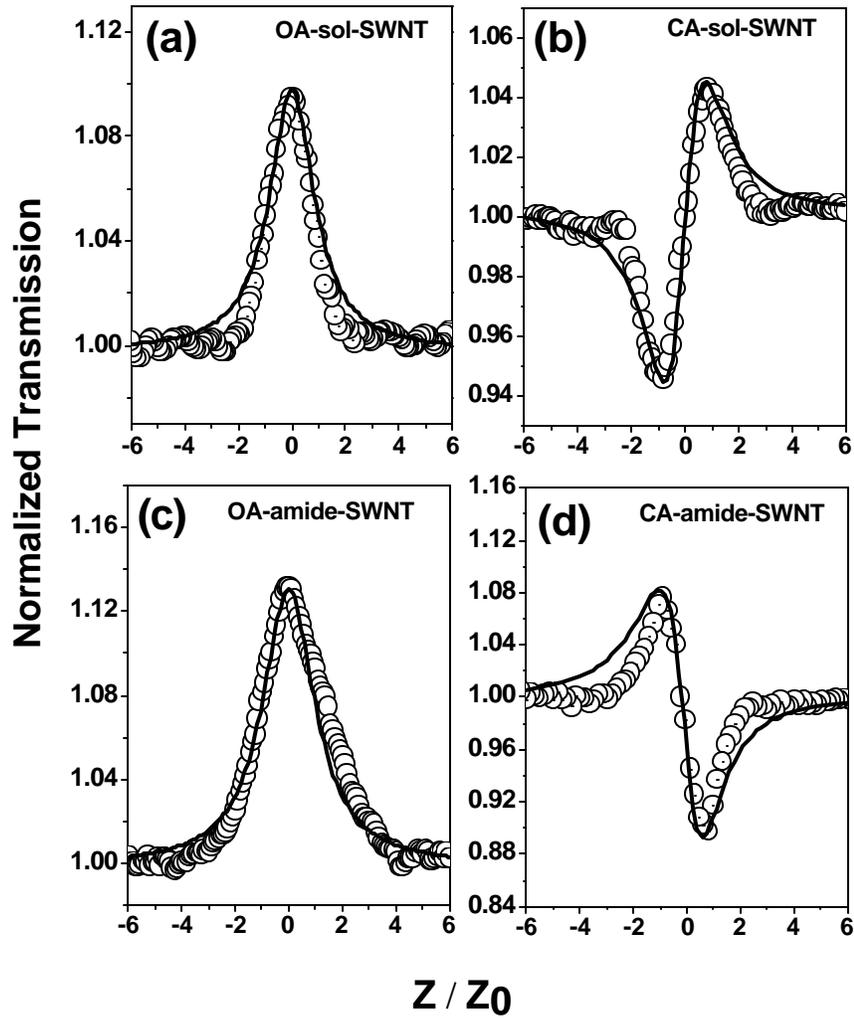

Fig. 3: Normalized transmittance data (open circles) in (a) and (b) are for sol-SWNT whereas (c) and (d) are for amide-SWNT. (a) and (c) are for OA z-scan case and (b) and (d) are for CA zscan case. Theoretical fit are shown by solid lines, giving parameters listed in Table 1.

Kamaraju et al.

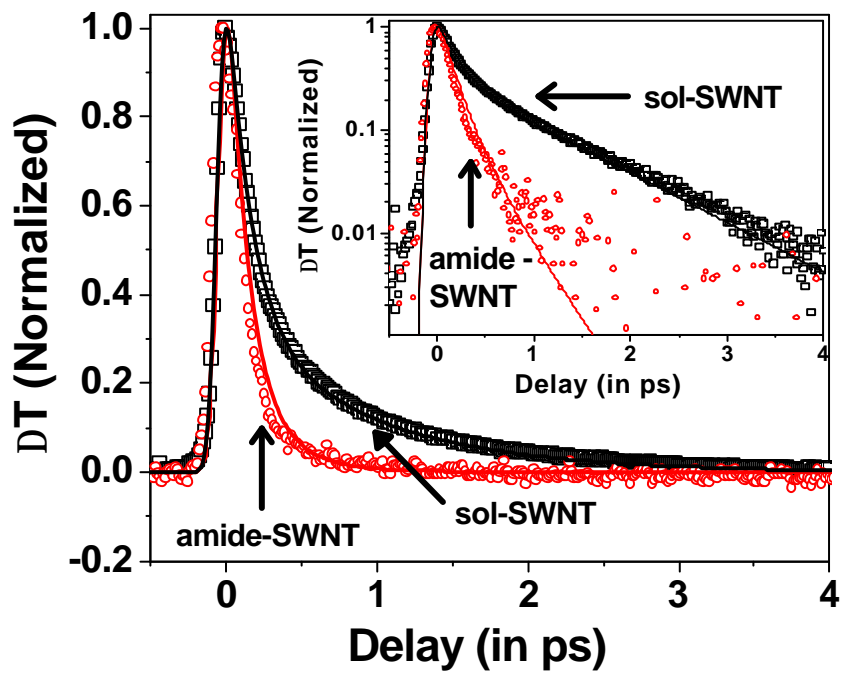

Fig. 4: Normalized differential transmittance vs delay between the pump and probe pulses for sol-SWNT (black hollow squares) and amide-SWNT (red hollow circles). The black (red) solid line is the convolution of the cross correlation at the sample point and a bi-exponential ultrafast photo-bleaching of 160 fs (130 fs) and 920 fs (300 fs) for sol (amide)-SWNT. The inset shows the log plot of the two decays.

Kamaraju et al.